\documentclass[aps,pra,preprint]{revtex4}

\usepackage{dcolumn}
\usepackage{graphicx} 
\usepackage{longtable}
\usepackage{amsmath,amsfonts,bm}
\usepackage{subfigure}
\usepackage{epstopdf}

\usepackage{array}
\newcolumntype{R}[1]{>{\raggedleft  \arraybackslash}p{#1}@{} }
\newcolumntype{C}[1]{>{\centering \arraybackslash}p{#1}@{} }

\usepackage{amsmath, amsfonts}

\newcommand{\be}{\begin{equation}}
\newcommand{\ee}{\end{equation}}
\newcommand{\bea}{\begin{eqnarray}}
\newcommand{\eea}{\end{eqnarray}}
\newcommand{\eq}[1]{\begin{align}#1\end{align}}
\newcommand{\Eq}[1]{Eq.\,(\ref{#1})}

\newcommand{\Ref}[1]{Ref.~\onlinecite{#1}}

\begin{document}

\title{  Using quadrature and an iterative eigensolver to compute 
 fine-structure ro-vibrational
    levels of Van der Waals complexes:
NH($^3\Sigma^-$)-He,
O$_2$($^3\Sigma^-_g$)-Ar and
O$_2$($^3\Sigma^-_g$)-He
}

\author{Xiao-Gang Wang}
\email{
xgwang.dalian@gmail.com}

\author{Tucker Carrington Jr.}
\email{Tucker.Carrington@queensu.ca}

\affiliation{%
Chemistry Department, Queen's University,
Kingston, Ontario
          K7L 3N6,
          Canada }

\date{Apr 19, 2018; Dec 18, 2018}

\begin{abstract}  %

We introduce a new method for computing spectra of molecules for which a spin-spin term in the Hamiltonian has an important effect.   In previous calculations, matrix
elements of the spin-spin term and of the potential were obtained by expanding the potential and using analytic equations in terms of $ 3-j $ symbols.   
Instead, we use quadrature.   Quadrature is simple and makes it possible to do calculations with a general potential and without using the Wigner-Eckart theorem.     In previous calculations, the Hamiltonian matrix was built and diagonalized.  Instead, we use an iterative eigensolver.  It makes it easy to work with a large basis.    The ideas are tested by computing energy levels of   NH($^3\Sigma^-$)-He,
O$_2$($^3\Sigma^-_g$)-Ar and
O$_2$($^3\Sigma^-_g$)-He.

\end{abstract}

\maketitle

\newpage

\section{Introduction}

Multi-dimensional quadrature makes it possible to compute the ro-vibrational spectrum of a molecule, even when the potential energy surface (PES)  does not have a special form that facilitates calculating integrals. 
Individual matrix elements could themselves  be computed by using multi-dimensional quadrature, however, by  using an iterative eigensolver in conjunction with multi-dimensional quadrature, spectra can be computed without calculating potential matrix elements. 
\cite{perspec,dvrrev,h3,pranab,lef,guo,our-h2o2,nno,our-h2o-h2}    
 It is only necessary to evaluate 
matrix-vector products (MVPs).  A MVP transforms an input vector into an output vector.   The output vector can be obtained by doing sums sequentially.  
This sequential summation procedure has important advantages:  1) it can be used regardless of the functional form of the PES; 2) the cost of 
calculating MVPs scales as it would if the PES were a sum of products (SOP), i.e., it is not more expensive to use a general PES than a SOP PES  (and  is less expensive unless the SOP has very few terms );  3) there is no need to derive and use closed-form equations for complicated integrals.  
MVPs are evaluated efficiently by exploiting product structure imposed by using a basis of products of 1-D functions and a product quadrature grid. 

In this paper, we show that  a quadrature  method for computing   the  ro-vibrational spectrum of a molecule for which interaction between rotational
angular momenta and electronic spin angular momentum is important has the same advantages.   
We are not the first to include this  interaction in a variational calculation. \cite{Tennyson-o2-ar,Tennyson-o2-he,Avoird83jcp}   
Several  effective fitting Hamiltonians that include such interactions have been derived and used.   \cite{waf1,waf2,waf3,oohf,oonno}  
In previous variational calculations, elements of a Hamiltonian matrix were computed using  closed-form expressions and 
 a Hamiltonian matrix was stored   and  diagonalized with    a direct
(not iterative) algorithm, whose cost scales as $N^3$, where $ N $  is the size of the matrix.   To do this one relies   on a special representation of the PES in terms of Legendre polynomials and  uses formulae involving 3$-j $ 
symbols and reduced matrix elements to calculate matrix elements of both the PES and the  spin interaction term.  The method we present works with general PESs.   We obviate the need for 3-$j $
symbols and concerns about phase conventions etc. by computing matrix elements of the spin interaction term and the PES   with quadrature.   All MVPs are done by 
evaluating sums sequentially.  In this paper, we propose equations for a complex with three atoms AB-R;
AB is a diatomic molecule in a $^3 \Sigma$ electronic state with electron spin  $\bm{S} = 1$ and  
R is a closed-shell atom with zero spin.
Everything is done using a body-fixed basis that is advantageous  when the anisotropy of the PES is strong.  We test our equations by computing energy levels for 
O$_2$-Ar, O$_2$-He and  NH-He.   Tennyson and co-workers were the first to do calculations on O$_2$-Ar, O$_2$-He.  \cite{Tennyson-o2-ar,Tennyson-o2-he} 
 Van der Avoird and Brocks\cite{o2-dimer}  have done calculations for  O$_2$ dimer.

\section{Vibration-rotation-spin kinetic energy operator}

\begin{figure}[!ht]
\begin{center}
\mbox{
\subfigure{\includegraphics[scale=0.4,angle=0]{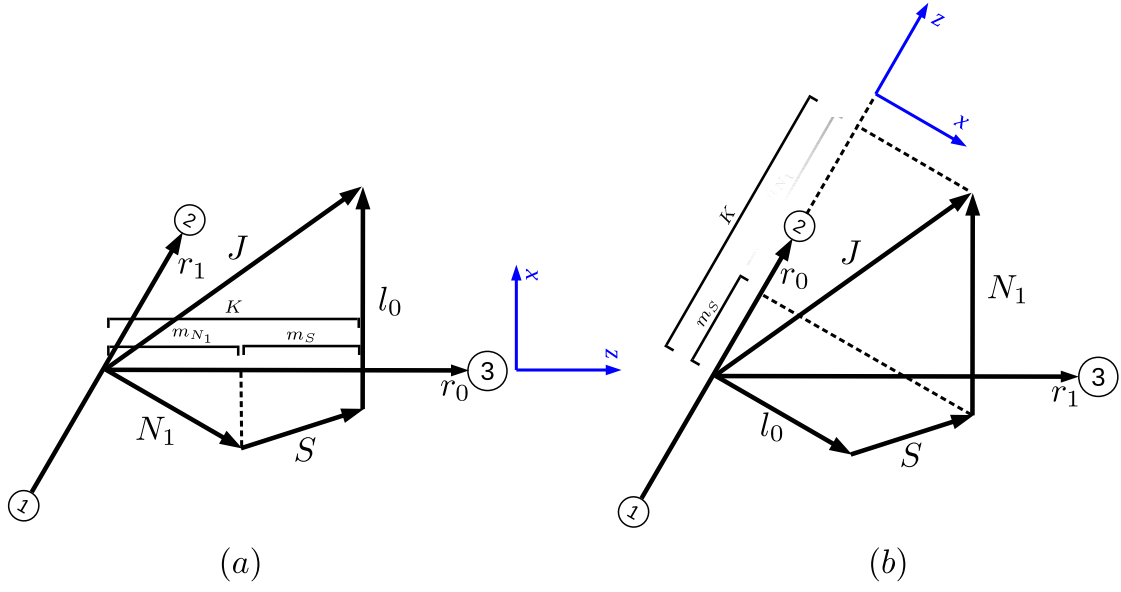}}
}
\caption{
Two coupling schemes used for systems consisting of an open shell $\Sigma$ diatomic molecule and a closed shell atom, such as O$_2$-Ar or NH-Ar.
In scheme (a),  the vector $r_0$ is associated with the intermonomer  Jacobi vector.
In scheme (b), the  vector $r_0$ is associated with the diatomic vector.
The $ z  $  axes of the body-fixed (BF)   (marked in blue)   and dimer-fixed  frames     in both schemes   are along  $\bm {r_0}$
and the $x$-axes  of the BF frames are  along  $\bm{ r_0 \times r_1 \times r_0}$. }
\label{fig_frame}
\end{center}
\end{figure}

To derive the correct kinetic energy operator (KEO), it is best to begin with the   dimer-fixed (DF) frame  KEO for a complex composed of two monomers which we denote
A and B.   
The DF frame is a  two-angle body-fixed (BF) frame.  It is  obtained by rotating a space-fixed (SF) frame by two Euler angles $(\alpha, \beta)$  so that 
 the DF frame z-axis is aligned with $\bm{r_0}$, the vector  connecting the centers of mass  of 
 A and B.  See Fig. \ref{fig_frame}a. 
As shown by Brocks {\em et al.}\cite{Brocks83}, the DF KEO is
\eq{
T^{\rm DF}
&= T^A + T^B + T_{int}
\nonumber \\
T_{int} &=
- \frac{1}{2\mu_0 r_0^2} \frac{\partial}{\partial r_0} r_0^2\frac{\partial}{\partial r_0}
+
B_{0}(r_0) \Bigl[
  J^2 -  \cot \beta \frac{\partial }{\partial \beta}+ \bm{j}^2 - 2 \bm{j} \cdot \bm{J}
           \Bigr]~,
\label{bast}
}
where  $ B_{0}(0)  =  1 / (2 \mu_{r_0} r_0^2)$.
 $\mu_{0}$ is the reduced mass.
$\bm{J}$ is the total angular momentum.  
$\bm{j}$ is the sum of  the  angular momenta of the two monomers.  
$T^A$ and $T^B$ are  KEOs  for  monomers  A and B in the DF frame.   Components of 
$\bm{J}$ and $\bm{j}$  are in the DF frame. 
The components of $\bm{J}$ have a special form 
and are written in terms of $\alpha$, $\beta$ and $j_z$ (See  Eq. (20) of Brocks {\em et al.}\cite{Brocks83}).

In this paper, we consider systems in which  monomer  A is a diatomic molecule in a 
$^3 \Sigma^-_g$ electronic state with electron spin  $\bm{S} = 1$ and monomer  B is an atom.
The total angular momentum of  monomer  A is 
\eq{
 \bm{j} = \bm{N_1} + \bm{S} ~,
\label{eq_j}
}
where  $\bm{N_1}$ is the rotational angular momentum  associated with  $\bm{r_1}$, the diatomic vector. See Fig. \ref{fig_frame}a. 
$ \bm{N_1} $ is usually called $\bm{l_1}$, if there is no spin.
In the DF frame, the spherical polar angles for vector $\bm{r_1}$ 
are $(\theta_1,  \gamma)$ 
( $\gamma$ is used for the azimuthal angle to be consistent with the  standard three-angle BF frame 
whose orientation in the SF frame is specified  by three Euler angles $(\alpha, \beta, \gamma)$).
The expressions for  $\bm{N_1}$ operators  in terms of  $( \theta_1, \gamma)$ are standard, 
see e.g.  Eq. (1.36) of \Ref{Zare88book}.
The spin-rovibrational KEO for monomer A is\cite{Mizu75} 
\eq{
T^A  =  
- \frac{1}{2\mu_1 r_1^2} \frac{\partial}{\partial r_1} r_1^2\frac{\partial}{\partial r_1}
+  B_1(r_1) \bm{N_1}^2
+ T_{\rm fine}  ~.
\label{T_A}
}
It is the usual diatomic KEO plus   effective spin-spin and spin-rotation interaction terms\cite{Mizu75},
\be
T_{\rm fine} = \frac{2}{3} \lambda_0 (3 S_\zeta^2 - S^2) + \gamma_0 \bm{ N_1 \cdot S} ~,
\label{tfine1}
\ee
where $S_\zeta$ is the projection of $\bm{S}$ along the diatomic vector.
$T^B = 0$ because B  is an atom.
Substituting \Eq{eq_j} and \Eq{T_A} into \Eq{bast},  we obtain
\eq{
T^{\rm DF}
= T_{\rm str} + 
B_{0}(r_0) \Bigl[
  J^2 -  \cot \beta \frac{\partial }{\partial \beta}+ (\bm{N_1 + S})^2 - 2 (\bm{N_1 + S}) \cdot \bm{J}
           \Bigr]
+  B_1(r_1) \bm{N_1}^2   +  T_{\rm fine}  ~,  %
\label{keo}
}
where
\eq{
T_{\rm str} =
- \frac{1}{2\mu_0 r_0^2} \frac{\partial}{\partial r_0} r_0^2\frac{\partial}{\partial r_0}
- \frac{1}{2\mu_1 r_1^2} \frac{\partial}{\partial r_1} r_1^2\frac{\partial}{\partial r_1}~.
\label{stre}
}
In this paper, 
the diatomic is rigid and therefore the second  term in \Eq{stre} is removed.    
In summary, to derive the KEO that includes   coupling between spin and rotation, 
one replaces $\bm{j}$ in  $T_{int}$    with $\bm{N_1 + S}$, but uses  $\bm{N_1}$  and not $\bm{N_1 + S}$ in 
  $T^A$.  
     $T_{\rm fine} $ $also$ couples  spin and rotation.

\section{Basis and matrix elements}

Knowing the KEO and having a potential, we calculate vibration-rotation-spin energy levels by choosing a basis and evaluating matrix-vector products (MVPs) to use the Lanczos algorithm.  When spin effects are omitted this is a common procedure. \cite{dvrrev,mjb,yum,buda,guo}    
When the Lanczos vectors are not orthogonalized, the memory and CPU costs are low.   To calculate wavefunctions we repeat the Lanczos iteration.\cite{cw,mjb,methane9d}   %
The  vibration-rotation-spin basis is a direct product of a stretch basis and a bend-rotation-spin basis.   The stretch basis functions  are discrete variable representation (DVR) functions. \cite{dvrrev}
In the rest of the section,  we  discuss only  the bend-rotation-spin basis.
Each bend-rotation-spin basis function is a  product,
\eq{
| N_1 m_{N_1} ; J K M ;  S m_S  \rangle  %
=
Y_{N_1}^{m_{N_1}} (\theta_1, \gamma)
\sqrt{\frac{2J+1}{4 \pi}} 
D^{J}_{M K}(\alpha, \beta, 0)^\ast
| S m_S \rangle ~,
\label{basis}
}
with   
$ K  \equiv  m_{N_1}  + m_S  $.
$m_{N_1}$, $m_S$, and $K$ are projection quantum numbers, along the dimer-fixed (DF) z-axis, 
 of the  rotational angular momentum $\bm{N_1}$, the electronic spin 
$\bm{S}$,
and the total angular momentum $\bm{J}$.   
Note that for a triatomic molecule with no spin, we normally use 
$ | l_1 m_{1} \rangle $ instead of   $ | N_1 m_{N_1} \rangle $ for the angular basis associated with  $\bm{r_1}$.
In this paper, we choose to use $ | N_1 m_{N_1} \rangle $ because we include the electronic spin angular momentum for the diatomic.
The basis of \Eq{basis} can be written in terms of a  standard  symmetric top Wigner function, 
\eq{
| N_1 m_{N_1} ; J K M ;  S m_S  \rangle  
=
\Theta_{N_1}^{m_{N_1}} (\theta_1)
\sqrt{\frac{2J+1}{8 \pi^2}} 
D^{J}_{M K}(\alpha, \beta, \gamma)^\ast
e^{-i m_S \gamma}
| S m_S \rangle ~,
\label{basis2}
}
because $ K  \equiv  m_{N_1}  + m_S  $.
If $S = 0$, \Eq{basis2}  is  the standard triatomic vibration-rotation basis.    \cite{tenny}
Tennyson and Mettes\cite{Tennyson-o2-ar} did  the first variational calculation for O$_2$-Ar. They used     the SF frame basis,
$ | ((N S) j L ) J M_J \rangle$,  
 in the notation of \Ref{Tennyson-o2-ar}, with $M_J$ the SF Z-axis projection of total $J$.
Their basis was built by successively coupling pairs of  angular momenta with Clebsch-Gordan coefficients.
In our notation, this basis would be 
$ | ((N_1 S) j l_0 ) J M \rangle$.
A notable difference between their basis and  our BF basis, \Eq{basis}, is that the BF basis does not have the $l_0$ quantum number
because  the   $z$-axis of the BF frame is along    $\bm{r_0}$.   The BF basis also does not have the $j$ label.  It does have  projection quantum numbers along the  BF z-axis.
To analytically evaluate potential integrals in their  SF basis,   Tennyson and Mettes    expand the  potential in terms of Legendre polynomials.  In the SF basis, equations for the  matrix elements of    $T_{\rm fine}$  are known in terms of     3-$j$ and 6-$j$ symbols\cite{Mizu75}.   They are derived with the  Wigner-Eckart theorem.  

We are not the first to propose using the BF basis of \Eq{basis}. 
Van der Avoird used it in  \Ref{Avoird83jcp}, with  perturbation theory, to explain the variational results of 
Tennyson and Mettes\cite{Tennyson-o2-ar}. Using the BF projection quantum numbers $K$ and $m_{N_1}$, Van der Avoird identified  ladder patterns  
in the variational levels. Using the Wigner-Eckart theorem\cite{Mizu75}, he   derived analytic  expressions for matrix elements of 
  $T_{\rm fine}$  in the BF basis. 
 These expressions involve a 3-$j$ symbol and double-bar reduced matrix elements. In section IV, we will propose a new method for  calculating matrix elements of  $T_{\rm fine}$ in the BF basis. 
 Van der Avoird and Brocks\cite{o2-dimer} use a BF basis to  variationally compute fine-structure levels of  O$_2$ dimer, a problem  more challenging than the triatomic in this paper.   
As mentioned above, operators for the components of  $\bm{J}$   in the DF frame  have a special form. Nonetheless, if %
$m_{N_1}$ in  \Eq{basis2} is equal to $K-m_S$, one obtains
\eq{
\hat{j}_z | N_1 m_{N_1} ; J K M ;  S m_S  \rangle  
&= (\hat{N}_{1z} + \hat{S}_z ) | N_1 m_{N_1} ; J K M ;  S m_S  \rangle  
\nonumber \\
&= (m_{N_1} + m_S ) | N_1 m_{N_1} ; J K M ;  S m_S  \rangle   
\nonumber \\
&= K | N_1 m_{N_1} ; J K M ;  S m_S  \rangle   ~.
}
Because of this equation, the  matrix elements of  $\bm{J}$ operators  in the DF frame  can be obtained from the usual standard equations, 
as in the case of systems with no spin.\cite{naf}  
$J^2 - \cot \beta {\partial }/{\partial \beta}$   
is diagonal and its non-zero elements are   $ J(J+1) $.      Matrix elements of $ J_{\pm}  $ are also equal to the usual expressions.  \cite{Zare88book}
The constraint  
 $ K = m_{N_1}+ m_S$ 
has a physical interpretation. As we see from Fig. \ref{fig_frame}a, 
the constraint arises because
$\bm{J = l_0 + N_1 + S}$ and  $\bm{l}_0$ has no projection on the z-axis of the DF frame.

Non-zero diagonal KEO matrix elements, in the basis of \Eq{basis},   for all terms except 
 the $S_{\zeta}^2$ term, which we shall focus on later in the paper,  
are:
\eq{
& \langle N_1 m_{N_1};J K M;S m_S|
T^{\rm DF}    %
| N_1 m_{N_1} ; J K M ;  S m_S  \rangle
\nonumber \\
=&  B_0 
\Bigl[
J(J+1) + N_1(N_1+1) + S(S+1) + 2 m_{N_1} m_S - 2K^2
\Bigr]
\nonumber \\
&- \frac{2}{3} \lambda_0 S(S+1)
+ \gamma_0 m_{N_1} m_S ~,
\label{element1} }
where the last two terms are  from $T_{\rm fine}$.
KEO terms  that are  off-diagonal are 
$ \bm{N_1 }  \cdot \bm{J}$,
$ \bm{S }  \cdot \bm{J}$,
and
$ \bm{N_1 }  \cdot \bm{S}$ from $T_{int}$
and 
$ \bm{N_1 }  \cdot \bm{S}$ from $T_{\rm fine}$.  Their matrix elements are computed from the relations, 
\eq{
 \bm{N_1 }  \cdot \bm{J}
&= \frac{1}{2} 
\bigl(
N_{1+} J_{-}  + N_{1-} J_{+} 
+ 2 N_{1z} J_{z}    \bigr)
\nonumber \\
 \bm{S }  \cdot \bm{J}
&= \frac{1}{2} 
\bigl(
S_{+} J_{-}  + S_{-} J_{+} 
+ 2 S_{z} J_{z}    \bigr)
\nonumber \\
 \bm{N_1 }  \cdot \bm{S}
&= \frac{1}{2} 
\bigl(
N_{1+} S_{-}  + N_{1-} S_{+} 
+ 2 N_{1z} S_{z}    \bigr) ~,
}
where 
$N_{1\pm} = N_{1x} \pm i N_{1y}$,
$J_{\pm} = J_{x} \pm i J_{y}$,
and
$S_{\pm} = S_{x} \pm i S_{y}$.
Excluding the $S_{\zeta}^2$ term, the non-zero off-diagonal matrix elements  are
\eq{
N_{1+} J_-:  \quad &
 \langle N_1, m_{N_1}+1;J,K+1,M;S m_S|
T^{\rm DF}
| N_1 m_{N_1} ; J K M ;  S m_S  \rangle
 = -B_0 \lambda^+_{N_1,m_{N_1}} \lambda^+_{JK}
\nonumber \\
S_{+} J_-:  \quad &
 \langle N_1, m_{N_1};J,K+1,M;S m_S+1|  
T^{\rm DF}
| N_1 m_{N_1} ; J K M ;  S m_S  \rangle
 = (-B_0) \lambda^+_{S,m_S} \lambda^+_{JK}
\nonumber \\
N_{1-} S_+: \quad &
\langle N_1, m_{N_1}-1;J,K,M;S m_S +1|
T^{\rm DF}
| N_1 m_{N_1} ; J K M ;  S m_S  \rangle
 = (B_0 + \frac{1}{2}\gamma_0) \lambda^-_{N_1,m_{N_1}} \lambda^+_{S m_S}~;
\label{element2}}
the  operator responsible for the off-diagonal element  is  given at  the beginning of each line and 
$\lambda^\pm_{JK} = \sqrt{J(J+1) - K(K \pm 1)}$ etc.

\subsection{Parity-adapted basis}
We prefer to work with a parity-adapted uncoupled basis which has the advantage of being two times smaller.
   The  parity-adapted basis functions are  equal to or  linear combinations of the uncoupled basis functions,
\bea
&&u_{N_1 m_{N_1} K m_S }^{JMSP} =
N_{ K m_S } \frac{1}{\sqrt{2}}   \times   %
\nonumber \\[1ex]
&&\left[
| N_1 m_{N_1} ; J K M \rangle | S m_S  \rangle  
+ (-1)^{J+S+P}
| N_1 \bar{m}_{N_1} ; J \bar{K} M \rangle | S \bar{m}_S  \rangle  
\right]~,
\label{pa_basis}
\eea
with
\[
N_{ K m_S } = (1 +   \delta_{K,0} \delta_{m_S ,0} )^{-1/2}~,
\] where $P=0$ and 1 correspond to even and odd parity, respectively.
The parity adapted basis functions are obtained by considering,   
\be
E^\ast | N_1 m_{N_1} ; J K M \rangle | S m_S  \rangle  
= (-1)^{J+S}| N_1 \bar{m}_{N_1} ; J \bar{K} M \rangle | S \bar{m}_S  \rangle  ~,
\label{Estar}
\ee
which is obtained from  $E^\ast | S m_S  \rangle   = (-1)^S | S \bar{m}_S  \rangle  $ (See Eq. (A18) of \Ref{Zare73}) 
and 
$ E^\ast | N_1 m_{N_1} ; J K M \rangle = (-1)^J | N_1 \bar{m}_{N_1} ; J \bar{K} M \rangle $ (See e.g. Eq. 10 of \Ref{nk3}).
Note that the $^3\Sigma^-$ electronic wavefunction has odd parity. This factor is not included in \Eq{Estar}  %
and not included in our tables and figures.  
In this work, we label levels  with  their spectroscopic parity $e/f$, which is   $(-1)^{J+P} = \pm 1$, respectively. 
For a single parity,  the number of parity-adapted functions is roughly half  the number of non-parity-adapted basis functions.  
The parity-adapted and non-parity-adapted bases span the same space if
the quantum numbers are restricted by
\bea
m_S  & \ge & 0,
\nonumber \\
K &\ge& 0 \quad ({\rm if~}m_S =0)~.
\label{m_constraint}
\eea
These restrictions complicate the evaluation of MVPs in the parity-adapted basis and we therefore transform from the PA basis to the basis of 
\Eq{basis2}, evaluate the MVP in that basis, and then transform back to the PA basis.
Recall that $ m_{N_1} $ is determined by $ m_S  $ and $ K $ and is not an independent basis label. %
Van der Avoird also used a  parity-adapted basis in Eq. 12 of \Ref{Avoird83jcp}. 
His notation $(-1)^p$ is  equivalent to our $(-1)^{J+S+P}$ in \Eq{pa_basis}.     
This is confirmed by comparing the assignments of our  O$_2$-Ar levels,  reported  in section VIII,   
with those of  \Ref{Avoird83jcp}.

\section{Matrix elements of the spin interaction term $S_\zeta^2$} 

We  use the basis of \Eq{basis2} because
 when the  anisotropy   of the PES is strong  %
 it is an excellent basis.    
Unfortunately, it is not simple to apply the spin interaction term to one of the basis functions.  The problem is that the spin interaction term is written in terms of $S_\zeta$, 
the component of the spin operator along the diatomic axis,   and the basis label $m_S$ is  the  quantum number for the  $ z $ component of the spin operator along the DF $ z  $ axis which is inter-monomer Jacobi vector.   
In order to calculate matrix elements of the spin interaction term, we write each of the basis functions in terms of functions of a basis in which $S^2_\zeta$ is diagonal.   
%
%
The relation between these two spin bases is\cite{Zare88book}
\be
| S m_S  \rangle
=
\sum_{\Sigma=-1}^{+1} D^{1 \ast}_{m_S \Sigma} (\gamma, \theta_1, 0)  
| S \Sigma \rangle 
=
e^{i m_S \gamma}
\sum_{\Sigma=-1}^{+1} d^{1}_{m_S \Sigma} (\theta_1)
| S \Sigma \rangle  ~,
\label{eq_transform}
\ee
where $ \Sigma $ is the component along the DF $ z $ axis.   
Now, a  $S^2_\zeta$ matrix element can be evaluated by doing  a sum,
\eq{
&\langle  N'_1 m'_{N_1} ; J K' M ; S m'_S | 
S_\zeta^2 
| N_1 m_{N_1} ; J K M  ; S m_S  \rangle  
\nonumber \\
& =
\sum_{\Sigma=-1}^{+1}
\Sigma^2
\langle  N'_1 m'_{N_1} ; J K' M ; S m'_S | 
e^{-i m'_S \gamma}
e^{i m_S \gamma}
d^{1}_{m'_S \Sigma} (\theta_1)
d^{1}_{m_S \Sigma} (\theta_1)
| N_1 m_{N_1} ; J K M  ; S m_S  \rangle   ~.
\label{s2_a}
}
The integral in \Eq{s2_a} is diagonal in $M$ because the operator in the middle does not depend on the Euler angle $\alpha$.  
We now show that the integral in  \Eq{s2_a} is also  diagonal in $K$.  In  \Eq{s2_a}, the integral over $\gamma$  is
\eq{
 \int_0^{2\pi} d \gamma 
\frac{1}{\sqrt{2\pi}} e^{-i( m'_{N_1} + m'_S)\gamma}
\frac{1}{\sqrt{2\pi}} e^{ i( m _{N_1} + m _S)\gamma}
= \delta_{K',K}~,
}
because  $ m_{N_1} + m_S = K $. Without the  factor $ e^{i m_S \gamma}$ in \Eq{eq_transform},  \Eq{s2_a}  would not be diagonal in $ K $.  %
Owing to the diagonality in $ K $,  the integrand of the  integral over $ \beta $ in 
 \Eq{s2_a}  is a product of two orthonormal  functions, 
$d^{J}_{M,K}(\beta) d^J_{M,K}(\beta)$  and is therefore unity.  %
We can  therefore re-write  \Eq{s2_a} as,
\eq{
&
\langle  N'_1 m'_{N_1} ; J K  M ; S m'_S | 
S_\zeta^2 
| N_1 m_{N_1} ; J K M  ; S m_S  \rangle  
\nonumber \\
&=
\sum_{\Sigma=-1}^{+1}
\Sigma^2
\int_{0}^{\pi} d \theta_1 \sin \theta_1
\Theta_{N'_1}^{m'_{N_1}} (\theta_1)
d^{1}_{m'_S \Sigma} (\theta_1)
d^{1}_{m_S \Sigma} (\theta_1)
\Theta_{N_1}^{m_{N_1}} (\theta_1)
\nonumber \\
&=
\sum_{\Sigma=-1}^{+1}
\Sigma^2
\sum_{\alpha_1}
T_{N'_1,\alpha_1}^{m'_{N_1}}
d^{1}_{m'_S \Sigma} (\theta_{\alpha_1})
d^{1}_{m_S  \Sigma} (\theta_{\alpha_1})
T_{N_1,\alpha_1}^{m_{N_1}}     \nonumber \\
&=
\sum_{\alpha_1}
T_{N'_1,\alpha_1}^{m'_{N_1}}
A_{m'_S m_S, \alpha_1}
T_{N_1,\alpha_1}^{m_{N_1}}~,
\label{szeta2_int}
} 
with
\be
A_{m'_S m_S, \alpha_1}
=
\sum_{\Sigma=-1}^{+1}
\Sigma^2
d^{1}_{m'_S \Sigma} (\theta_{\alpha_1})
d^{1}_{m_S  \Sigma} (\theta_{\alpha_1})
\label{A}
\ee
and
\be
T_{N_1,\alpha_1}^{m_{N_1}}
=
\sqrt{w_{\alpha_1}} \Theta_{N_1}^{m_{N_1}}(\theta_{\alpha_1}) ~,
\ee
where
$(w_{\alpha_1}, \theta_{\alpha_1}) $ are Gauss-Legendre quadrature weights and points.
$N_1$, $m_{N_1}$ and $m_S$ are coupled by the spin term. 
The sum over $ \alpha_1 $  in \Eq{szeta2_int}  has almost the same form as a standard quadrature approximation for a  matrix element of a 1-D $ \theta $  potential in an associated Legendre basis.  However in \Eq{szeta2_int},
   $ 
A_{m'_S m_S, \alpha_1}$   is off-diagonal in    $m_{S}  $.   
  For a particular  $ (m'_S, m_S)$ pair, $ 
A_{m'_S m_S, \alpha_1}$ 
 is a sum of terms $ \sin^{n_s}\theta   \cos^{n_c}\theta $, where  $n_c+n_s  \le 2 $. 
Despite the fact that there are terms with $ n_s=1 $, we use  Gauss Legendre quadrature.  It is nearly  exact when the number of points is chosen to be large enough to converge potential integrals.  
A  matrix-vector product with the $S_\zeta^2$ matrix in \Eq{szeta2_int} can be evaluated by doing sums sequentially,
\be
x'_{N'_1 m'_{N_1},J K,S m'_S}
= \sum_{\alpha_1} T_{N'_1,\alpha_1}^{m'_{N_1}}
\sum_{m_S} A_{m'_S m_S, \alpha_1}
\sum_{N_1} T_{N_1,\alpha_1}^{m_{N_1}} x_{N_1 m_{N_1},J K,S m_S} ~,
\label{s2_mv}
\ee  %
where 
$x_{N_1 m_{N_1},J K,S m_S}$ is a vector in the basis of \Eq{basis}.
Because of the constraint,    $m_{N_1}$  on the input vector is not an independent label and is computed from  $m_{N_1} = K - m_S$, and 
  $m'_{N_1}$  on the output vector is not an independent label and is computed from  $m'_{N_1} = K - m'_S$.  

Rather than using \Eq{s2_mv}, one could  calculate the matrix elements on  the left side  of  \Eq{szeta2_int}, by  doing  the sum over $\alpha_1$.
   After  doing the sum over $\alpha_1$, one would obtain the same numbers that 
Van der Avoird gets using  his  analytic equation  in terms of  $ 3-j  $ symbols and  double bar reduced matrix elements.
His  equation shows that  $N_1$ is coupled only to  $ N'_1 = N +  \Delta N_1  $  with $\Delta N_1    \le 2$.
The memory cost of storing the non-zero elements of the $S_\zeta^2$ matrix therefore scales as $ N_{\rm bas}$,
where $N_{\rm bas}$ is the size of the basis.
One could  use  stored non-zero elements of the $S_\zeta^2$ matrix to compute the matrix-vector product required to use an  iterative eigensolver.
The cost of the MVP would then  scale as  $ N_{\rm bas}$.   
We prefer to store the small $\mathbf{A}$ matrix of \Eq{A} and  use sequential sums, as in \Eq{s2_mv}, to compute the matrix-vector product
For our choice, the memory cost is smaller and the cost of the matrix-vector product is similar. 
Not using analytic equations for matrix elements when it is possible might seem less elegant, but quadrature reduces the memory cost and  (not in this case but in general) reduces the cost of evaluating MVPs.   This is strikingly obvious for potential MVPs.   Quadrature is not approximate and does not increase the cost of the calculations
Moreover, the  quadrature  approach would  also  be straightforward  even if   $ T_{fine} $ involved complicated functions of 
 $S_\zeta$ operators and even if the angle between the $ \zeta $  axis and the BF $ z $  axis were a complicated function of the shape of the molecule.
It also has the advantage that one does not need  to  %
 be careful about phase factor errors in  analytical expressions.

\section{Assignment of quantum numbers}

$J$ and $ P $   
 are conserved quantum numbers and we calculate levels with different values separately.    It is fairly straightforward to assign values of  
 $K$, $m_S$, $m_{N_1}$, and $N_1$ to every energy level because the  basis functions have these labels. 
To assign values of $l_0$ and $j_0$ is harder because  the  basis functions do not have these labels.  
The assignments of projection quantum numbers $K$, $m_S$, $m_{N_1}$ are made with the approach of \Ref{our-h2o-h2}.  
For example,   $K$ is assigned by summing the squares of expansion coefficients in the non-PA basis
over all indices other than $K$ to obtain a quantity we call $P_K$, which  is a measure of the contribution of basis functions labelled by  $K$ to the full wavefunction.
( $\sum_K P_K = 1$.)
$ P_K $ and $ P_{-K} $ are equal   
and 
$ P_{m_{N_1}} $ and $ P_{-m_{N_1}} $ are equal.  
We therefore cannot  assign the signs of  $K$, $m_S$, and $m_{N_1}$, however, using the relation, $K = m_{N_1} + m_S$, 
we can  determine  whether   $m_S$, and $m_{N_1}$ have the same or different signs.  
For   O$_2$-Ar or O$_2$-He,  this allows  us to determine whether  the  DF $ z $   projections for   
the rotation of O$_2$ and  the spin of O$_2$ are aligned or anti-aligned. 

We also assign $l_0$, $j$ and $N_1$ by computing expectation values of $\bm{l_0^2}$, $\bm{j^2}$ and $\bm{N_1^2}$, denoted by 
$\langle \bm{ l_0^2} \rangle$, $\langle \bm{ j_0^2} \rangle$,  and $\langle \bm{ N_1^2} \rangle$, respectively,
for every state.  These are useful approximate good quantum numbers if  coupling is weak. 
For example,  the  expectation value of $\bm{N_1^2}$ in state $ \psi^i $ is   
\begin{eqnarray}
\langle \bm{ N_1^2} \rangle &=& \langle \psi^i \vert  {N_1^2} \vert  \psi^i \rangle  \nonumber \\
 &=& \sum_{N'_1,K',m'_S \atop  N_1,K,m_S   }  C^i_{N'_1,K',m'_S}  C^i_{N_1,K,m_S}    \langle  N'_1 m'_{N_1} ; J K M ;  S m'_S  
   \vert  {N_1^2} \vert 
     N_1 m_{N_1} ; J K M ;  S m_S  \rangle  
\nonumber \\     
 &=& \sum_{N_1,K,m_S   }    (C^i_{N_1,K,m_S})^2  N_1(N_1+1)   ~,
\end{eqnarray}
where  %
\begin{eqnarray}
\ \vert  \psi^i \rangle  
= \sum_{N_1,K,m_s}  C^i_{N_1,K,m_s}      | N_1 m_{N_1} ; J K M ;  S m_S  \rangle  ~.
\end{eqnarray}
The coefficients,  $  C^i_{N_1,K,m_s}   $ in the non-PA basis  can be easily obtained from the coefficients in the PA basis used to do the calculation.  
We use the non-PA basis because its off-diagonal elements are easier to deal with (in the PA basis they are  complicated due to the restrictions on the quantum numbers in \Eq{m_constraint}).    
Computing $\langle \bm{ N_1^2} \rangle$ is easy because the basis functions are labelled by $ N_1 $.
Computing $\langle \bm{ l_0^2} \rangle$  and $\langle \bm{ j^2} \rangle$     is  not as straightforward
 because   our basis functions do not have $  l_0 $ and $ j_0  $ labels.
A good way to calculate    $\langle \bm{ l_0^2} \rangle$  is to use 
 $\bm{ l_0^2} = \bm{ (J - N_1 - S) ^2}$.
The required matrix elements  are given in Eqs. \ref{element1} and \ref{element2}. 
For $\langle \bm{ j^2} \rangle$, we compute  the expectation values of $\bm{ j^2} = \bm{ ( N_1 + S) ^2}$.
Unlike $\bm{N_1^2}$,   off-diagonal matrix elements  contribute to the sums  for $\bm{ l_0^2}$ and $\bm{ j^2}$.

\section{An alternative method of computing the spectrum}

 In the previous sections of this paper, we propose (method I) using quadrature with  a body-fixed KEO whose $ z $ axis is along the inter-monomer Jacobi vector from the centre of mass of O$_2$ to the rare gas atom, in order to treat 
 the spin interaction term  proportional to $S_\zeta^2$. 
 In this section we outline a different method (method II) that does not require any quadrature and uses   a body-fixed KEO whose $ z $ axis is along the  diatomic O$_2$ vector.  This alternative method has disadvantages, but can be used to check calculations.   
 In method I, quadrature is necessary because the spin basis used to compute the spectrum, $ \vert S m_S \rangle  $,  has the label $ m_S $ for the projection along the inter-monomer Jacobi vector  and must therefore be replaced with a linear combination of spin functions,  $ \vert S \Sigma \rangle  $,    labelled by $\Sigma  $ (\Eq{eq_transform}),  
 the projection along the diatomic Jacobi vector, in which 
  $S_\zeta^2$ is diagonal.
In method II,  the inter-monomer Jacobi vector is  $\bm{{r_1}}$ and the diatomic Jacobi vector is  $\bm{{r_0}}$.
In our notation, the BF $ z  $  axis is  along  $\bm{r_0}$.
%
%
The obvious advantage of method II is    that $S_\zeta^2$ is  diagonal, i.e.  $ S_\zeta^2 | S m_S \rangle =  m_S^2 | S m_S \rangle$  :  there is no need to introduce a second set of spin functions.

An important  disadvantage of method II  is that the Coriolis coupling is much larger than for method I.  
Take NH-He as an example, the rotational constants are:  $b_{\rm NH}$ = 16.343 cm$^{-1}$ and $b_{\rm inter}$ = 0.475 cm$^{-1}$   
(both evaluated  at the equilibrium geometry on the PES used in this work).
In method I, $B_0$ = $b_{\rm inter}$ = 0.475 cm$^{-1}$ and in method II, $B_0$ = $b_{\rm NH}$ = 16.343 cm$^{-1}$.   It is clear that 
 the Coriolis interaction term, proportional to $B_0$, is much larger in method II.  
A second and perhaps more critical disadvantage of method II  is that it is difficult to  apply it  to larger systems such as O$_2$-CO where the second monomer is a closed-shell spinless diatomic molecule 
rather than an atom.   This is because if the BF  $ z $ is along the O$_2$ vector then  one needs 
 polyspherical angles  for CO that are angles relating  the  orientation of  two unconnected monomers and in terms of these angles 
  potential coupling is large.   %

%

The KEO in method II has the same form as the KEO in method I.  However,  $r_0$ is now the diatomic Jacobi vector and $ r_1  $ is now the inter-monomer Jacobi vector. 
The basis functions are  formally identical to those of method I, \Eq{basis} and \Eq{basis2}.  Although  %
 $\bm{{r}_1}$ is now the inter-monomer  Jacobi vector  and using  $\bm{N_1}$ for its    angular momentum is unconventional, we prefer not to change the notation.
The KEO terms $T^{\rm DF}$,  $T_{int}$ and $T^A$ take the same form.
The only formal change in the KEO is in the spin-rotation interaction term, the $\gamma_0$-related term (cf. \Eq{tfine1}),
\be
T_{\rm fine} = \frac{2}{3} \lambda_0 (3 S_\zeta^2 - S^2) + \gamma_0 \bm{ (J - N_1 - S) \cdot S} ~.
\label{tfine2}
\ee
This change  is necessary  because in method I 
the rotation of the diatomic molecule with spin is described by  $\bm{r_1}$ and its angular momentum $\bm{N_1}$
and in method II the rotation of the diatomic molecule with spin is described by  $\bm{{r}_0}$ and its angular momentum $\bm{l_0 = J - N_1 - S}$.
Because of this change, the diagonal and off-diagonal matrix elements change slightly.
Non-zero diagonal KEO matrix elements, in the basis of \Eq{basis},   for all terms {\em including}
 the $S_{\zeta}^2$ term, 
are:
\eq{
& \langle N_1 m_{N_1};J K M;S m_S|
T^{\rm DF}
| N_1 m_{N_1} ; J K M ;  S m_S  \rangle
\nonumber \\
=&  B_0 
\Bigl[
J(J+1) + N_1(N_1+1) + S(S+1) + 2 m_{N_1} m_S - 2K^2
\Bigr]
\nonumber \\
&
+ 2 \lambda_0 m_S^2
- \frac{2}{3} \lambda_0 S(S+1)
- \gamma_0 \Bigl[ S(S+1) + m_S^2 \Bigr] ~.
\label{element1x} }
Compared to \Eq{element1}, the new term $ 2 \lambda_0 m_S^2$ is from  the $  2 \lambda_0 S_{\zeta}^2$ term of $T_{\rm fine}$
and the $\gamma_0$-related  term  has changed.
\Eq{element1} does not have the $  2 \lambda_0 S_{\zeta}^2$ term.


%
Including the $S_{\zeta}^2$ term, the non-zero off-diagonal matrix elements  are
\eq{
N_{1+} J_-:  \quad &
 \langle N_1, m_{N_1}+1;J,K+1,M;S m_S|
T^{\rm DF}
| N_1 m_{N_1} ; J K M ;  S m_S  \rangle
 = -B_0 \lambda^+_{N_1,m_{N_1}} \lambda^+_{JK}
\nonumber \\
S_{+} J_-:  \quad &
 \langle N_1, m_{N_1};J,K+1,M;S m_S+1|
T^{\rm DF}
| N_1 m_{N_1} ; J K M ;  S m_S  \rangle
 = - (B_0 - \frac{1}{2}\gamma_0) \lambda^+_{S,m_S} \lambda^+_{JK}
\nonumber \\
N_{1-} S_+: \quad &
\langle N_1, m_{N_1}-1;J,K,M;S m_S +1|
T^{\rm DF}
| N_1 m_{N_1} ; J K M ;  S m_S  \rangle
 = (B_0 - \frac{1}{2}\gamma_0) \lambda^-_{N_1,m_{N_1}} \lambda^+_{S m_S}~,
\label{element2x}}
with the corresponding operator given at the beginning of each line. 
Compared to  \Eq{element2}, the second and third equations are different due to the change of  the $\gamma_0$-related term in the KEO.

\section{Results: Spin-rovibrational levels of NH($^3\Sigma^-$)-He}

Using both method I and method II, we computed the  spin-rovibration levels of NH($^3\Sigma^-$)-He,
which were  previously reported  in \Ref{Cybulski05}.   
Cybulski et al. also used two methods:   the SF approach of  Tennyson and Mettes\cite{Tennyson-o2-ar,Tennyson-o2-he} and 
a BF method with the same basis we have in method I (the $ z $ angular momentum  and spin components that label basis functions are along the inter-monomer Jacobi vector).   The two sets of energies computed by Cybulski et al. differ by less than 
 0.0001 cm$^{-1}$.   Their accurate calculations 
allow  us to test our methods. 
We use the PES ``potential I'' of \Ref{Cybulski05} and  the same parameters:  
$b_{\rm NH} = 16.343 $ cm$^{-1}$ is the rotational constant of NH in its ground vibrational state;
$\lambda_0 = 0.920 $ cm$^{-1}$ and $\gamma_0 = -0.055 $ cm$^{-1}$ are, respectively, the spin-spin and spin-rotation interaction constants for NH.
 We also use the same atomic masses, as in \Ref{Cybulski05}. The NH distance is fixed throughout the calculations.

The size of the  bend basis is determined  by max($N_1$) = 25; Cybulski et al. used a maximum value of 8.
 30 Gauss-Legendre quadrature points are used for $\theta_1$, to integrate the potential 
and $S_{\zeta}^2$ term in Eq. 19. 
The stretch basis for $r_0$ is 120 sine DVR functions in the range [3.5, 35] $a_0$.
The total basis size is small enough that when using    an iterative eigensolver, there is no need to    optimise   the stretch basis.

Cybulski et al.  calculate levels with and without     $T_{\rm fine}$.   The two sets of levels 
are labelled  ``no spin'' and ``with spin'' in Table \ref{tab1}.  
``No spin'' levels with $ l_0 >  0 $ are split into three closely spaced "with spin" levels.  
For their   ``no spin'' levels  Cybulski et al. use the dissociation limit  on their PES  as the  zero of energy.  For their ``with spin'' levels they use a different zero of energy.  The zero of energy for  ``with spin'' is the 
experimental\cite{Bernath86jms}   energy of the lowest rotational-spin NH($^3\Sigma^-$) level.     It is  0.0077 cm$^{-1}$\cite{Cybulski05,Bernath86jms} below the  
dissociation energy on the PES.  
This 0.0077 cm$^{-1}$ shift is due to  $T_{\rm fine}$. 
Our levels are compared to those of  Cybulski et al.   in  Table \ref{tab1}.  For both our ``no spin'' and  our ``with spin" levels, the zero of energy is the dissociation limit.
To compare  Cybulski et al.'s and   our ``with spin'' results, it is therefore necessary to add -0.0077 cm$^{-1}$  to the 
with spin  
energies in  \Ref{Cybulski05}.

There is nearly perfect agreement between our levels and those of \Ref{Cybulski05}.
Most differences between our energy levels and their counterparts in \Ref{Cybulski05}  are less than  0.0001 cm$^{-1}$ which is the stated accuracy of \Ref{Cybulski05}.   
The  three 
  levels with $l_0=3$ have larger differences.       
 Both with and without the spin term,  our levels are lower by about 0.0014 cm$^{-1}$.  Gonz\'alez-Mart\'inez and Hutson\cite{Hutson07pra} computed the same levels with   the BOUND program\cite{BOUND} 
 and also obtained levels  about  0.0014 cm$^{-1}$ lower than those of \Ref{Cybulski05}.    They attribute these  differences to the large  $r_0$   tails of the corresponding  wavefunctions.    Our calculation supports    this observation since we find it is necessary to use a larger $max(r_0)$ = 35 $a_0$ rather than 30 $a_0$ of \Ref{Cybulski05} to converge these three levels.
Assigned values of $ l_0 $ are also given in Table \ref{tab1}.   The  $ l_0 $  values are    obtained from 
 $\langle \bm{ l_0^2} \rangle$ and are close to    $  l_0 (l_0 + 1)$. 
Our $l_0$ assignments agree with those of \Ref{Cybulski05}; they use   $ L $ rather than  $l_0$.
Likewise, our  $J$ and parity assignments in  Table \ref{tab1}  agree with those of \Ref{Cybulski05}.
We also give   $K$, $m_S$ and $m_{N_1}$ assignments that are not in  \Ref{Cybulski05}.
Most of  the reported  bound states  can be assigned to  $K=0$ or  1.  However,  some are  a mixture of $K=0$ and 1 and the corresponding $ K $ values are indicated as   $0/1$ or $1/0$, with the first number being the one with the  largest contribution to the wavefunction.
 Because the  NH rotational constant is large,   all  the states have  $N_1$ = 0.   
 All calculated $\langle \bm{ N_1^2} \rangle$ are smaller than 0.005.
 NH($^3\Sigma^-$)-He has a  shallow potential well ($D_e$ = 19.84 cm$^{-1}$)  and states with    $N_1 > 0 $ are not bound.  %
   Due to the constraint, $K = m_S + m_{N_1}$, $m_S$ assignments are the same as $K$ assignments and therefore $m_S$ assignments are not given in
Table \ref{tab1}.
Finally, $\langle \bm{ j^2} \rangle$ ranges from 2.002 to 2.005, indicating that all levels are  $j{\rm (NH)}=1$ states.  $j{\rm (NH)}$ is a good quantum number because spin is strongly coupled to NH rotation. 
As  the potential well is shallow and the barrier to NH rotation is low (3 to 4 cm$^{-1}$)\cite{Cybulski05}, 
we expect  $ j $ to be a good quantum number.
Finally, NH($^3\Sigma^-$)-He  energy levels computed with  method II (section VI ) and the same basis sizes  agree with those computed with method I  to 9 decimal places (in cm$^{-1}$).    The basis is evidently large enough that the levels are well converged despite  the larger Coriolis coupling of method II.
When using an iterative eigensolver, basis size is not a problem.   %
Since the two methods deal with $S_\zeta^2$ term  differently, this excellent   agreement indicates 
that both methods are correct.  

\section{Results: Spin-rovibrational levels of O$_2$($^3\Sigma^-_g$)-Ar }

As  mentioned previously, Tennyson and Mettes\cite{Tennyson-o2-ar} computed  O$_2$($^3\Sigma^-_g$)-Ar levels with a variational method.  
In this paper, we use a  PES more recent than the one used by  Tennyson and Mettes.  
The new PES is obtained using  supermolecular unrestricted Moller-Plesset perturbation theory\cite{o2-ar-pes}.  
This computer code for this PES was kindly provided to us by Mark Severson and is deposited in the supplementary material of this paper.   
At equilibrium the molecule is T-shaped with $\theta_{1e} = 90 ^\circ$ and $r_{0e} = 6.7 a_0$\cite{o2-ar-pes}; the 
 well depth is 117 cm$^{-1}$.
We use  atomic masses; for the  rotational constant of O$_2$ in its ground vibrational state\cite{o2-const} we use 
$b_{\rm O_2} = 1.437678 $ cm$^{-1}$; $\lambda_0 = 1.98475 $ cm$^{-1}$ and $\gamma_0 = -0.00845 $ cm$^{-1}$ are, respectively, the spin-spin and spin-rotation interaction constants for O$_2$\cite{o2-const}.
The bend basis size  is defined by max($N_1$) = 25. Only odd rotational states of O$_2$ are physically allowed and we therefore include only   basis functions with odd $N_1$ values. 
30 Gauss-Legendre quadrature points are used for $\theta_1$, to integrate the potential 
and in Eq. 19.  
The stretch basis for $r_0$ is 120 sine DVR \cite{dvrrev} functions in the range [3.0, 35] $a_0$.
The O$_2$ distance is fixed throughout the calculation as in the PES.\cite{o2-ar-pes} %

To understand the pattern of the  O$_2$($^3\Sigma^-_g$)-Ar levels, it is useful to discuss  the 
 O$_2$ levels.
The $N_1=1$ O$_2$ rotational levels 
are  split   by the spin  interaction term $T_{\rm fine}$ in \Eq{tfine1}
into three levels.
These three levels are from low to high, 
$j_{ {\rm O}_2} = 0, 2 $ and 1 levels, and are between 0 and  4 cm$^{-1}$, determined by the value of $\lambda_0$.
Here $\bm{j_{ {\rm O}_2}}$ is the  total angular momentum of ${\rm O}_2$.
%
%
%
However, due to the anisotropy of the PES, $j_{{\rm O}_2} $ is not a good quantum number in O$_2$($^3\Sigma^-_g$)-Ar.
Van der Avoird beautifully explained the results  of \Ref{Tennyson-o2-ar} by using a simple model.   \cite{Avoird83jcp}   
Starting from the BF basis \Eq{basis} and using  perturbation theory, 
he noted  that the positive coefficient $V_2(r_0)$  of the  $ P_2(\cos\theta)$ term in the potential expansion, which is responsible for 
 the T-shaped  structure of the O$_2$-X complex, pushes 
the  $m_{N_1} = 0$ levels up relative to the $m_{N_1} = \pm 1$ levels by $\frac{3}{5} V_2(r_0) \approx 12 $ cm$^{-1}$.
This is much larger than the magnitude of the splittings in  free O$_2$ caused  by the spin  interaction term $T_{\rm fine}$. 
Here $m_{N_1} $ refers to the component of $N_1$ along  the BF z axis. 
Four lower-energy "ladders" are built on six 
zeroth-order states obtained from products of 
 the two $m_{N_1} = \pm 1$ basis functions  with three spin function with $m_S = 0, \pm 1$.
 The rungs of two of the ladders are doublets.  The two components of each doublet have different parity. 
  Two of the ladders have $K=0$, one has $K=1$, and one has $K=2$. 
These four ladders are labelled $i=1,2,3$ and 4.\cite{Avoird83jcp}\
Two higher-energy ladders are built on three 
zeroth-order  states   that are  products of 
the $m_{N_1} = 0$ basis function and  the    three spin function with $m_S = 0, \pm 1$.  
One  of these two ladders has  
 $K=0$  and the other has  $K=1$.  
The rungs of  the $K=1$ ladder are doublets. 
These two ladders are labelled $i=5$ and 6.\cite{Avoird83jcp}
No variational calculation for ladders $i=5$ and $i=6$ have been reported for  O$_2$($^3\Sigma^-_g$)-Ar, 
but they have been  reported for O$_2$($^3\Sigma^-_g$)-He.\cite{Tennyson-o2-he}
The characteristics of all  six ladders are summarized in Table \ref{ladder_character}.

\begin{figure}[!ht]
\begin{center}
\includegraphics[scale=0.6,angle=0]{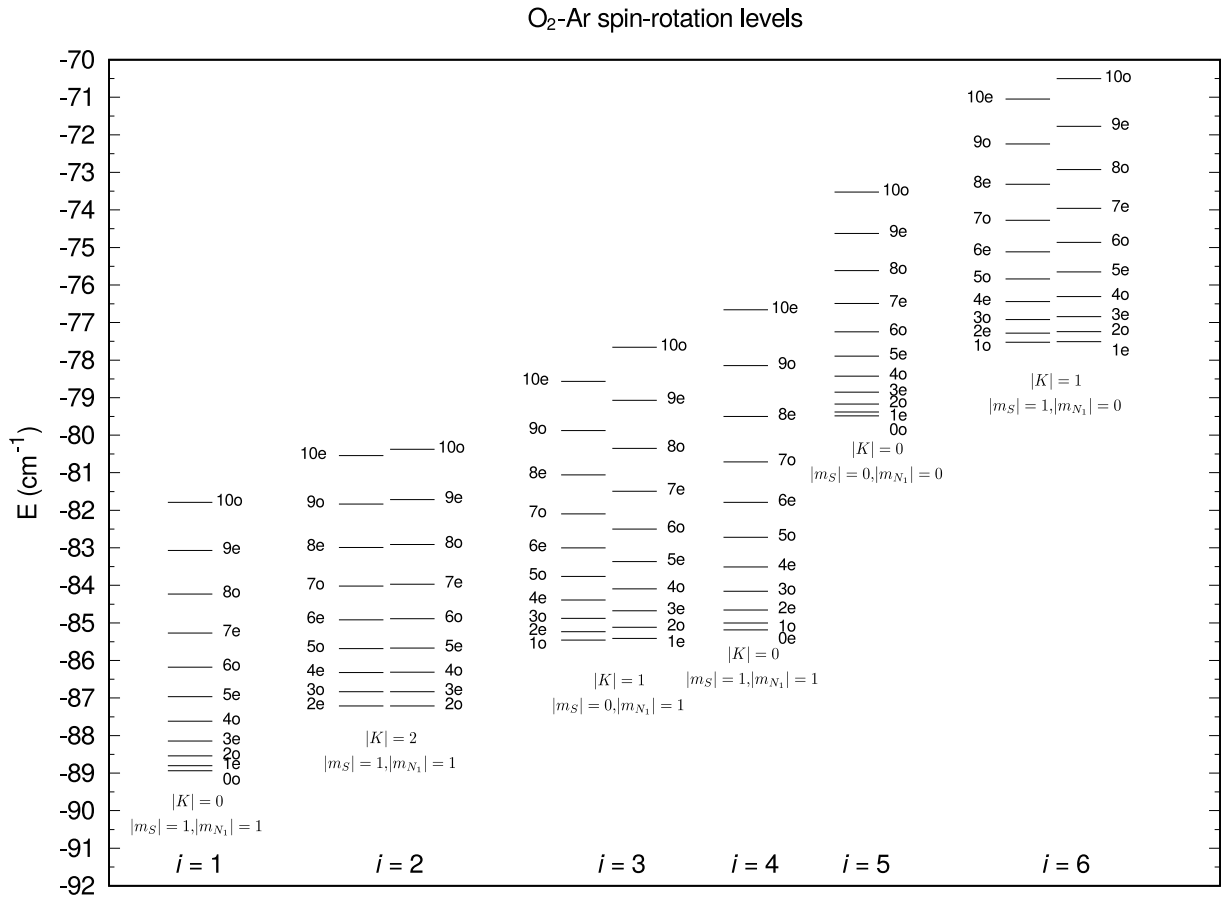}
\caption{
Fine-structure rovibrational levels of O$_2$-Ar.
Theoretical levels are sorted into ladders based on the approximate quantum numbers 
$K$, $m_S$, and $m_{N_1}$. %
The first four ladders  $i=1, \cdots, 4$ originate from 
$m_{N_1}=1$ states of O$_2$.
The next two ladders  $i=5, 6$ originate from 
$m_{N_1}=0$ states of O$_2$.
Indicated with  each level is $J(e/o)$ where $e/o$ is even/odd parity.
}
\label{fig2}
\end{center}
\end{figure}

Using wavefunctions, we  assign $K$, $m_S$ and $m_{N_1}$ values  to each level. These labels make it possible to sort the levels we compute into 
Van der Avoird's ladders.
See Table \ref{levels1} for the  ladders of assigned levels  and Fig. \ref{fig2} for an illustrative level diagram.
Our four lower ladders including their  $K$, $m_S$, $m_{N_1}$, and parity assignments  agree well with those of   
\Ref{Tennyson-o2-ar}, shown in Fig. 1 of \Ref{Avoird83jcp} (most of the variational levels were not published)
even though the potential energy surfaces  are different.  The order of the even and odd doublets for the  $K>0$ ladders order also agrees.
The computed 
 $K$, $m_S$, $m_{N_1}$, 
 $\langle \bm{ l_0^2} \rangle$,  
 $\langle \bm{ N_1^2} \rangle$ 
and
 $\langle \bm{ j^2} \rangle$ values 
 for each level are given in the Supplementary Material (SM).   
As discussed above, due to the  barrier to O$_2$ internal rotation, coupling between  $N_1$   and  $\bm{ l_0}$ is more important than coupling between 
 $S$ and  $N_1$, and as a consequence neither  the  total rotation angular momentum $j$ of the diatomic nor the end-over-end rotation angular momentum $\bm{ l_0}$ are
 useful labels.
$\langle \bm{ l_0^2} \rangle$ values are not close to   $l_0 (l_0 + 1)$ , as they would be if  $l_0$ were a good quantum number, however,    they do  increase as one moves up the ladder.   
In contrast, $l_0$ is almost a good quantum number for  NH-He because it is closer to the free rotor limit.
Because the next O$_2$ rotational level $N_1=3$ is much higher in energy, all states shown are expected to have large contributions from  $N_1=1$ basis functions.
However,  $\langle \bm{ N_1^2} \rangle$ fluctuates between 2.7 and 4.3, far  from expected value of $2$.

The ladders of O$_2$-Ar  were recently used to analyse the level structure of    CO-O$_2$.\cite{co-o2}   %
In \Ref{co-o2}, CO-O$_2$ energy levels were divided into two groups 
and    a  $K=0$ ladder (referred to as a stack in \Ref{co-o2}) in group 1 and a $K=2$ ladder  in  group 2  were linked  with the
 $i=1$  ladder ($K=0$)  and the  $i=2$  ladder  ($K=2$) of O$_2$-Ar.   
 Groups and ladders are labelled by $n({\rm O}_2), j({\rm O}_2)$.  However, Van der Avoird\cite{Avoird83jcp} clearly established that $   
  j({\rm O}_2) $ is not a useful label for  O$_2$-Ar.   It seems likely that  $   
  j({\rm O}_2) $ is also not a useful label for   CO-O$_2$.

\section{Results: Spin-rovibrational levels of  O$_2$($^3\Sigma^-_g$)-He}

We use an accurate PES determined   with the  partially spin-restricted open-shell single and double excitation
coupled cluster method with perturbative triples.\cite{o2-he-pes}
This is a 3-D potential, but we fix the O$_2$ distance at the equilibrium value, $r_{1e} = 2.282 a_0$.
The potential well depth is 27.9 cm$^{-1}$ and the equilibrium geometry is  T-shaped  with $\theta_{1e} = 90 ^\circ$ and $r_{0e} = 6.00 a_0$\cite{o2-he-pes}.
Compared to O$_2$-Ar, the O$_2$-He potential well depth is much smaller and the rare gas mass is much lighter.
Therefore,  O$_2$-He states are more de-localized   
and the Coriolis coupling has a bigger effect on energy levels,   which makes it more difficult to organize the levels into ladders.   
The Coriolis coupling is larger for  O$_2$($^3\Sigma^-_g$)-He  because $B_0$ (see \Eq{bast}) is larger. 
We use  atomic masses and the same O$_2$ parameters as for  O$_2$($^3\Sigma^-_g$)-Ar.
The bend  and  stretch bases are the same as for  O$_2$($^3\Sigma^-_g$)-Ar.

All of our calculated O$_2$($^3\Sigma^-_g$)-He bound states are given in Table \ref{level2}.
The  ZPE is -7.47 cm$^{-1}$, it sits  about a quarter of the way up the well.    In the same table, we also list the levels computed by 
Tennyson and Van der Avoird\cite{Tennyson-o2-he} (TA) using a  variational method\cite{Tennyson-o2-ar} and an older empirical PES\cite{o2-he-pes-old}.
The PES used by  TA is certainly less accurate and  its  well  depth is only  23.5 cm$^{-1}$.  On their PES, 
TA found   35 bound states and on the newer PES we find only 28 bound states, even though the potential we use has a deeper well. 
Compared to O$_2$-Ar,
the  Coriolis coupling  has a greater effect on the energy levels.   Many of the     O$_2$-He    states have more than one dominant $K$ component. 
Moreover, a lot of states have more than one dominant $m_{N_1}$. See Table \ref{level2}.
Because the ladders are defined by $K$ and $m_{N_1}$, the mixing of these quantum numbers means that it is harder to identify states with ladders.

\begin{figure}[!ht]
\begin{center}
\includegraphics[scale=0.6,angle=0]{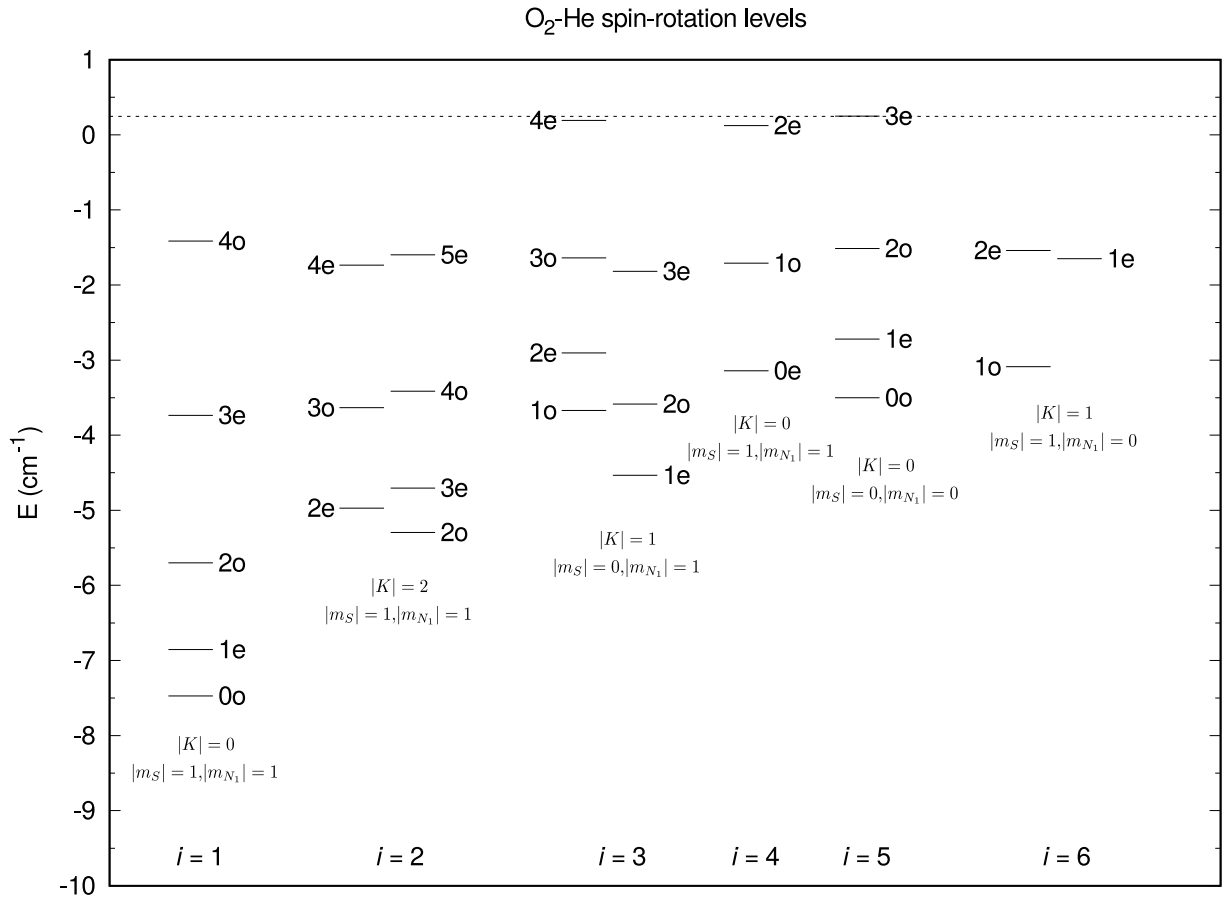}
\caption{
Fine-structure rovibrational levels of O$_2$-He. See the caption of Fig. \ref{fig2} for additional information about the ladders.
The dashed line is the dissociation limit with O$_2$ in its $J=1$ rotational state with energy 0.246 cm$^{-1}$.
}
\label{fig3}
\end{center}
\end{figure}

Tennyson and Van der Avoird\cite{Tennyson-o2-he} assigned ladder labels to their calculated levels by comparing with the results of a perturbation model, but 
noted that the  O$_2$-He ladders were  less regular than those of O$_2$-Ar.
They observed, for the first time,  two higher ladders $i=5$ and $i=6$ with $m_{N_1} = 0$ which they did not assign for O$_2$-Ar.
Using the newer PES, we are also able to identify  $i=5$ and $i=6$  ladders and find that all the ladders are harder to distinguish than for O$_2$-Ar.
 This is due to the stronger Coriolis coupling.
The simplest way to assign a ladder label $ i $ to each of our levels is to  compare our level list, for a fixed $ J $ and fixed parity, to the assigned 
level list of TA and use their assignments for our levels.  When this is done we observe that gaps between levels in a ladder are sometimes irregular.  
This means that the two PESs are different enough that we cannot always use the assignments of TA and we therefore  use 
 $m_{N_1}$, $K$, $\langle \bm{ j_{\rm O_2}^2} \rangle$ and  $\langle \bm{ l_0^2} \rangle$ to 
replace some of the TA labels attached to our levels.   Once this is done the gaps between the rungs of the ladders become regular.    
See Fig. \ref{fig3}. These re-assigned levels are marked by a star in Table \ref{level2}.
The reassignments exchange labels 
 between ladders $i=4$ and $i=6$ and between ladders $i=1$ and $i=2$.
 One type of reassignment is based on values of  $m_{N_1}$. The  $i=4$ and $i=6$ ladders have $m_{N_1} =$ 1 and 0, respectively. Thus, we reassign the the $J=1$ odd state at -3.087 cm$^{-1}$ which has  $m_{N_1}$ = 0  to the   $i=6$ ladder.   
 Another type of reassignment is based on values of 
  $K $,  $\langle \bm{ j_{\rm O_2}^2} \rangle$, and  $\langle \bm{ l_0^2} \rangle$.
  The rotation of  O$_2$    is less hindered in O$_2$-He  than in 
  O$_2$-Ar and therefore
   $\bm{j = N_1 + S}$, is more nearly  a good quantum number.         
   One can therefore  
   tentatively associate ladders $i=2,3,4$ with $j_{\rm O_2}$ = 2 (5 components),  ladders $i=5, 6$ with $j_{\rm O_2}$ = 1 (3 components), and    
ladder $i=1$  with $j_{\rm O_2}$ = 0 (one component).     
The ladders then have the same energy order as the states of 
O$_2$  for which the rotational energy order is $j_{\rm O_2}$ = 0, 2 and 1.
Another example of states that are reassigned is the two $J=3$ even states at -4.703 and -3.736 cm$^{-1}$.  They  are re-assigned to the  $i=2$ and $i=1$ ladders, respectively,  for three reasons.  
First, their   $\langle \bm{ j_{\rm O_2}^2} \rangle$  are 5.90 and 0.38, indicating $j_{\rm O_2} = 2$ and 0. 
Second,  their     $\langle \bm{ l_0^2} \rangle$  are 2.57 and 11.73, indicating $l_{0} = 1$ and 3.  
If ladder $i=1$ has $j_{\rm O_2} = 0$ then,  because  $\bm{J = j_{\rm O_2} + l_0 }$, 
  $J = l_0$ and therefore a $ J=3  $  level must have  $l_0=3$  and hence belong to ladder $i=1$. 
Third, the $K$ values of the two levels are  1/2/0 
and 0 which also supports this assignment because ladders $i=2$ and $i=1$ have $K=2$ and 1.
This example also shows that due to the  rotation of $O_2$ being less hindered, $l_0$ and $j_{\rm O_2}$ are close to being good quantum numbers for O$_2$-He, whereas they are basically useless for O$_2$-Ar,  as discussed in the previous section.

One serious problem in  the ladder assignment occurs for the $i=5$ ladder. The first $i=5$ state at -3.500 cm$^{-1}$ is a pure $K=0$ state as expected, but 
 the next three $i=5$ states are all pure $K=1$ states.  %
 States with different values of $ K $ should not occur in the same ladder.
 An illustrative level diagram for all the bound states of O$_2$($^3\Sigma^-_g$)-He is shown in Fig. \ref{fig3}. 
One can compare it with the same diagram in \Ref{Tennyson-o2-he} and see that the ladder pattern is quite different.
From Fig. \ref{fig3}, we see that level spacings between different ladders are rather different, indicating different values of end-over-end rotation constants. 
For all $K>0$ ladders, we find that the splitting between doublets of each rung is rather large, sometimes larger than the J-spacings within the ladder, which is in stark contrast with the O$_2$-Ar case.

\section{Conclusion}

In this paper we present a new approach for   including the spin-spin term that couples rotational angular momenta and electronic spin angular momentum.     It is new in two ways.   First, it is new because it uses an iterative eigensolver to solve a matrix representation of the 
Schroedinger equation.   Second, it is new because it does not expand the potential in terms of Legendre polynomials and use  equations involving $ 3-j $ symbols to obtain matrix elements of  both the potential and the spin-spin term.  Instead, quadrature is used for both and matrix-vector products are done by evaluating sums sequentially.    The method of this paper can be used without expanding the potential and without employing angular momentum theory to derive matrix elements.   This makes it easy to use with a general PES.   

One might think that using  quadrature when it is possible to replace matrix elements with exact closed-form equations would increase the 
cost of the calculation.  This is not true.   It is not true for the same reason that   the cost of the  MVPs  one must evaluate 
when using quadrature for PES matrix elements  and a SOP PES is the same.   In some cases we expect the new method to be cheaper.  
It is certainly cheaper if, when expanded in terms of Legendre polynomials, the PES has many terms and also cheaper for molecules 
for which the potential matrix is large.   Moreover,  if the molecule of the  molecule-rare gas complex were larger than a diatomic 
and if its shape was not fixed, then the standard approach of building and diagonalizing a matrix would be much more costly.    
Experiments have been done on many O$_2$ containing complexes, e.g.  N$_2$O-O$_2$, H$_2$O-O$_2$, HF-O$_2$ etc.   
The method of this paper opens the door to doing calculations on these complexes.

\section*{Acknowledgements}
The financial support of the Natural Sciences and Engineering Research 
Council is gratefully acknowledged.   
We thank Mark Severson for sending us the O$_2$-Ar PES published in \Ref{o2-ar-pes}.  %
%



\clearpage
\begingroup
\begin{table}[ht]
\centering
\caption{
Spin-rovibrational levels of NH($^3\Sigma^-$)-He (in cm$^{-1}$) relative to the dissociation energy with no $T_{fine}$ term.  The 
Cybulski columns are from \Ref{Cybulski05}. The NH electronic ground state energy -0.0077 cm$^{-1}$ (due to $T_{fine}$ ) was added to the data of  \Ref{Cybulski05} 
to obtain the spin-rovibrational levels in the second to last column. 
$e/o$ is even/odd parity.
 }
\begin{tabular}
{
 C{1.5cm}
 C{1.5cm}
 C{1.5cm}
 R{2.0cm}
 R{2.0cm}
C{0.5cm}
 R{2.0cm}
 R{2.0cm}
}
\hline \hline
&&&  \multicolumn{2}{c}{no spin}                 &&  \multicolumn{2}{c}{with spin}        \\
\cline{4-5}
\cline{7-8}
$J(e/o)$  &$l_0$      &    $K$        & Cybulski   & This work     && Cybulski   & This work    \\
\hline
 $1e $    & $0$       &     $1/0$       &  -4.4174   &  -4.4175      &&    -4.4251 & -4.4252     \\
 $0o $    & $1$       &     $0$         &  -3.7818   &  -3.7819      &&    -3.7867 & -3.7868     \\
 $1o $    & $1$       &     $1$         &            &               &&    -3.7909 & -3.7911     \\
 $2o $    & $1$       &     $1/0$       &            &               &&    -3.7892 & -3.7894     \\
 $1e $    & $2$       &     $0/1$       &  -2.5375   &  -2.5377      &&    -2.5442 & -2.5444     \\
 $2e $    & $2$       &     $1$         &            &               &&    -2.5462 & -2.5464     \\
 $3e $    & $2$       &     $1/0$       &            &               &&    -2.5449 & -2.5451     \\
 $2o $    & $3$       &     $0/1$       &  -0.7538   &  -0.7552      &&    -0.7613 & -0.7627     \\
 $3o $    & $3$       &     $1$         &            &               &&    -0.7619 & -0.7633     \\
 $4o $    & $3$       &     $1/0$       &            &               &&    -0.7614 & -0.7628     \\
\hline \hline
\end{tabular}
\label{tab1}
\end{table}
\endgroup

\clearpage
\begingroup
\begin{table}[ht]
\centering
\caption{
Labels  of ladders\cite{Avoird83jcp} for spin-rovibrational levels of O$_2$-Ar($^3\Sigma^-$).
$\sigma=(e,f)$ is the spectroscopic parity with $e$ and $f$ representing
$(-1)^{J+P} =$ +1 and -1, respectively.  %
 }
\begin{tabular}
{
 C{3cm}
 C{2.0cm}
 C{2.0cm}
 C{2.0cm}
 C{2.0cm}
 C{2.0cm}
 C{2.0cm}
}
\hline \hline
ladder     &  $K(\sigma)$ &  $m_S$  & $m_{ N_1 }$  & $N_1$  &  $(-1)^{J+S+P}$\\
\hline
 $i=1$     &   0$(f)  $  &   1     & 1      & 1     & $+   1$ \\
 $i=2$     &   2$(e,f)$  &   1     & 1      & 1     & $\pm 1$ \\
 $i=3$     &   1$(e,f)$  &   0     & 1      & 1     & $\pm 1$  \\
 $i=4$     &   0$(e)  $  &   1     & 1      & 1     & $-   1$  \\
 $i=5$     &   0$(f)  $  &   0     & 0      & 1     & $+   1$  \\
 $i=6$     &   1$(e,f)$  &   1     & 0      & 1     & $\pm 1$  \\
\hline \hline
\end{tabular}
\label{ladder_character}
\end{table}
\endgroup

\clearpage
\begingroup
\begin{table}[ht]
\centering
\caption{
Spin-rovibrational levels of O$_2$-Ar($^3\Sigma^-_g$) (in cm$^{-1}$) relative to the dissociation energy without $ T_{fine}$.  
We follow \Ref{Avoird83jcp} and  label the ladders  by $i$ in the first row. See also Table \ref{ladder_character} and  Fig. \ref{fig2}
for more information about the ladders.
 }
\begin{tabular}
{
 C{1.5cm}
 C{1.5cm} C{1.5cm} C{1.5cm} C{0.2cm} C{1.5cm} C{1.5cm} C{0.2cm}
 C{1.5cm} C{1.5cm} C{0.2cm} C{1.5cm} C{1.5cm} 
}
\hline \hline
         &  $i=1$   & \multicolumn{2}{c}{$i=2$}    && \multicolumn{2}{c}{$i=3$}    && $i=4$    & $i=5$    && \multicolumn{2}{c}{$i=6$}   \\
         \cline{3-4} \cline{6-7} \cline{12-13}
$J$	 & K=0f	    & K=2e     &  K=2f    &&  K=1e    &  K=1f    &&  K=0e    &  K=0f    &&  K=1e   &   K=1f   \\
\hline
0	 & -88.934  &          &          &&          &          && -85.182  & -79.484  &&         &          \\
1	 & -88.802  &          &          && -85.454  & -85.411  && -85.001  & -79.380  && -77.521 &  -77.507 \\
2	 & -88.538  & -87.212  & -87.212  && -85.231  & -85.117  && -84.656  & -79.170  && -77.28  &  -77.239 \\
3	 & -88.143  & -86.830  & -86.828  && -84.878  & -84.677  && -84.156  & -78.853  && -76.919 &  -76.839 \\
4	 & -87.617  & -86.321  & -86.313  && -84.389  & -84.093  && -83.510  & -78.428  && -76.438 &  -76.309 \\
5	 & -86.962  & -85.682  & -85.666  && -83.762  & -83.366  && -82.720  & -77.894  && -75.837 &  -75.651 \\
6	 & -86.178  & -84.915  & -84.885  && -82.998  & -82.499  && -81.787  & -77.248  && -75.116 &  -74.866 \\
7	 & -85.268  & -84.018  & -83.966  && -82.095  & -81.492  && -80.713  & -76.489  && -74.277 &  -73.958 \\
8	 & -84.231  & -82.991  & -82.910  && -81.055  & -80.348  && -79.499  & -75.616  && -73.318 &  -72.926 \\
9	 & -83.070  & -81.833  & -81.713  && -79.878  & -79.069  && -78.147  & -74.628  && -72.242 &  -71.774 \\
10	 & -81.786  & -80.545  & -80.377  && -78.565  & -77.656  && -76.656  & -73.524  && -71.047 &  -70.503 \\
\hline \hline
\end{tabular}
\label{levels1}
\end{table}
\endgroup

\clearpage
\squeezetable
\begingroup
\begin{table}[ht]
\centering
\caption{
All bound spin-rovibrational levels of O$_2$-He($^3\Sigma^-_g$) (in cm$^{-1}$) relative to the dissociation  energy without    $ T_{fine} $.
TA refers to Tennyson and van der Avoird\cite{Tennyson-o2-he}. TW is This Work.   To the TA levels of 
\Ref{Tennyson-o2-he}, we have   added  0.246 cm$^{-1}$, the energy of the $J=1$ O$_2$ state,  to account for the difference in the definition of the  zero
of energy in this paper and in TA.  
All levels in this table are below 0.246 cm$^{-1}$ and are bound, except for one level at 0.248 cm$^{-1}$.
 }
\begin{tabular}
{
 C{1.5cm} C{1.5cm}
 C{1.5cm} C{1.5cm} C{1.5cm}
 C{1.5cm} C{1.5cm} C{1.5cm}
 C{1.5cm} C{1.5cm} 
 }
\hline \hline
$E$    	  &$E$(TA) &$K$     &$m_S$   &$m_{N_1}$ 
& $\langle \bm{ l_0^2} \rangle$ & $\langle \bm{ N_1^2} \rangle$   & $\langle \bm{ j_{\rm O_2}^2} \rangle$ 
& ladder (TA) & ladder (TW)      \\
\hline
J=0,even  &        &        &        &        &        &        &        &           &        \\
-3.142	  &-5.515  &0       &1       &1       &2.16    &2.16    &2.16    &$i=4$      &$i=4$   \\
J=0,odd   &        &        &        &        &        &        &        &           &        \\
-7.471	  &-9.598  &0       &1       &1       &0.25    &2.15    &0.25    &$i=1$      &$i=1$   \\
-3.500	  &-4.484  &0       &0       &0       &5.81    &2.39    &5.81    &$i=5$      &$i=5$   \\
J=1,even  &        &        &        &        &        &        &        &           &        \\
-6.855	  &-9.027  &0       &1       &1       &2.19    &2.15    &0.29    &$i=1$      &$i=1$   \\
-4.534	  &-6.287  &1/0     &0       &0/1     &2.10    &2.26    &5.83    &$i=3$      &$i=3$   \\
-2.719	  &-4.262  &1       &1/0     &1/0     &2.46    &2.13    &2.36    &$i=5$      &$i=5$   \\
-1.650	  &-2.897  &0/1     &1/0     &0       &11.36   &2.29    &5.75    &$i=6$      &$i=6$   \\
J=1,odd   &        &        &        &        &        &        &        &           &        \\
-3.670	  &-6.032  &1       &0       &1       &1.94    &2.16    &3.34    &$i=3$      &$i=3$   \\
-3.087	  &-4.950  &1       &1       &0       &4.39    &2.15    &4.80    &$i=4$      &$i=6$*  \\
-1.708	  &-3.435  &0/1     &1       &1       &5.99    &2.12    &2.20    &$i=6$      &$i=4$*  \\
J=2,even  &        &        &        &        &        &        &        &           &        \\
-4.971	  &-7.303  &2       &1       &1       &2.20    &2.36    &6.13    &$i=2$      &$i=2$   \\
-2.904	  &-5.134  &1/0     &1/0     &1       &2.42    &2.14    &2.26    &$i=3$      &$i=3$   \\
-1.541	  &-3.520  &1       &1/0     &1/0     &11.72   &2.20    &5.94    &$i=4$      &$i=6$*  \\
0.122	  &-1.805  &0/1     &1       &1       &11.96   &2.11    &2.16    &$i=6$      &$i=4$*  \\
J=2,odd   &        &        &        &        &        &        &        &           &        \\
-5.701	  &-8.013  &0       &1       &1       &4.64    &2.21    &1.70    &$i=1$      &$i=1$   \\
-5.295	  &-7.342  &2/0/1   &1       &1       &1.83    &2.26    &4.65    &$i=2$      &$i=2$   \\
-3.584	  &-5.357  &2/0/1   &1/0     &1       &5.95    &2.33    &5.96    &$i=3$      &$i=3$   \\
-1.514	  &-3.198  &1       &1/0     &0/1     &6.09    &2.11    &2.20    &$i=5$      &$i=5$   \\
J=3,even  &        &        &        &        &        &        &        &           &        \\
-4.703	  &-7.303  &1/2/0   &1       &1       &2.57    &2.32    &5.90    &$i=1$      &$i=2$*  \\
-3.736	  &-5.134  &0       &1       &1       &11.73   &2.10    &0.38    &$i=2$      &$i=1$*  \\
-1.817	  &-3.520  &2/0     &1       &1       &11.98   &2.35    &6.04    &$i=3$      &$i=3$   \\
0.248	  &-1.805  &1       &1/0     &0/1     &11.93   &2.10    &2.19    &$i=5$      &$i=5$   \\
J=3,odd   &        &        &        &        &        &        &        &           &        \\
-3.635	  &-5.987  &2       &1       &1       &6.17    &2.33    &6.12    &$i=2$      &$i=2$   \\
-1.640	  &-3.814  &1/0     &1       &1       &6.20    &2.13    &2.17    &$i=3$      &$i=3$   \\
J=4,even  &        &        &        &        &        &        &        &           &        \\
-1.737	  &-4.128  &2       &1       &1       &12.12   &2.30    &6.11    &$i=2$      &$i=2$   \\
0.192	  &-1.990  &1/0     &1       &1       &12.12   &2.11    &2.14    &$i=3$      &$i=3$   \\
J=4,odd   &        &        &        &        &        &        &        &           &        \\
-3.414	  &-5.484  &1/0/2   &1       &1       &6.23    &2.30    &6.08    &$i=1$      &$i=2$*  \\
-1.415	  &-3.534  &0       &1       &1       &19.93   &2.09    &0.16    &$i=2$      &$i=1$*  \\
J=5,even  &        &        &        &        &        &        &        &           &        \\
-1.598	  &-3.628  &1/0     &1       &1       &12.13   &2.29    &6.09    &$i=1$      &$i=2$*  \\
\hline \hline                                                                           
\end{tabular}
\label{level2}
\end{table}
\endgroup

\clearpage

\clearpage

\clearpage

\end{document}